\documentclass[graybox]{svmult}

\usepackage{mathptmx}       
\usepackage{helvet}         
\usepackage{courier}        
\usepackage{type1cm}        
\usepackage{makeidx}         
\usepackage{graphicx}        
\usepackage{multicol}        
\usepackage[bottom]{footmisc}
\usepackage{amsmath}
\usepackage{amssymb}

\def\d{\text{d}}
\def\e{\text{e}}
\def\i{\text{i}}

\makeindex             

\begin{document}

\title*{Explicit formulae in probability and in statistical physics}
\author{Alain Comtet and Yves Tourigny}
\institute{Alain Comtet \at UPMC Univ. Paris 6, 75005 Paris and
  Univ. Paris Sud ; CNRS ; LPTMS, UMR 8626, \\ Orsay F-91405, France, \email{alain.comtet@u-psud.fr}
\and Yves Tourigny \at School of Mathematics, University of Bristol, Bristol BS8 1TW, United Kingdom, \\ \email{y.tourigny@bristol.ac.uk}}

%
%
\maketitle

\abstract{We consider two aspects of Marc Yor's work that have had an impact in statistical physics: firstly, his results on the windings of planar Brownian motion and their implications for the study of polymers; secondly,
his theory of exponential functionals of
L\'{e}vy processes and its connections with disordered systems. 
Particular emphasis is placed on techniques leading to explicit calculations.
}

\section{Introduction}
\label{introductionSection}
In this article, dedicated to Marc Yor, we would like to highlight some aspects of his work which have had a direct impact in statistical physics taken in its broader sense. Although Marc did not draw his inspiration from physics, he firmly believed in the unity of science and in the fruitfulness of approaching a problem from
many different angles. The seminar ``Physique et Probabilit\'es'' at the Institut Henri Poincar\'e, which he promoted, and in which he participated actively, attests his desire to share his work beyond a specialist circle. We were all struck 
by his fascination for explicit formulae. More than mere curiosities, explicit formulae can sometimes reveal deep connections between different probabilistic objects; in his hands, they generated further formulae and their inspection would often be illuminating. This attitude was in stark contrast with that adopted by certain mathematical physicists--- ``the austere guardians of the temple'' who often view their r\^{o}le as one of providing rigorous existence proofs. Without sacrificing rigour, Marc taught us to value the special case--- not only as a possible clue for the general case, but also for its own
intrinsic beauty.
In the following, we will discuss two among the many themes to which Marc made significant
contributions during his career: (1) the winding properties of planar Brownian motion and (2) exponential functionals of L\'evy processes. Our aim is to provide instances of application of his work to physical systems and to emphasize its relevance and innovative character.

\section{The winding number of planar Brownian motion}
\label{windingSection}

The subject takes off in 1958 with the pioneering work of  Spitzer \cite{Sp}, in which he proves
$$
\frac{2\theta_t}{\log t} \xrightarrow[t \rightarrow \infty]{\text{law}} C_1
$$
where  $Z_t=X_t+\text{i}Y_t$  is a planar Brownian motion, $\theta_t = \arg Z_t $ a continuous determination of the angle swept up to time $t$ and $C_1$ a Cauchy random variable of parameter $1$. Note that the same logarithmic scale is already  present in the work of Kallianpur and Robbins on occupation times \cite{KR}.
\begin{figure}[!ht]
  \centering
  \includegraphics[scale=0.30]{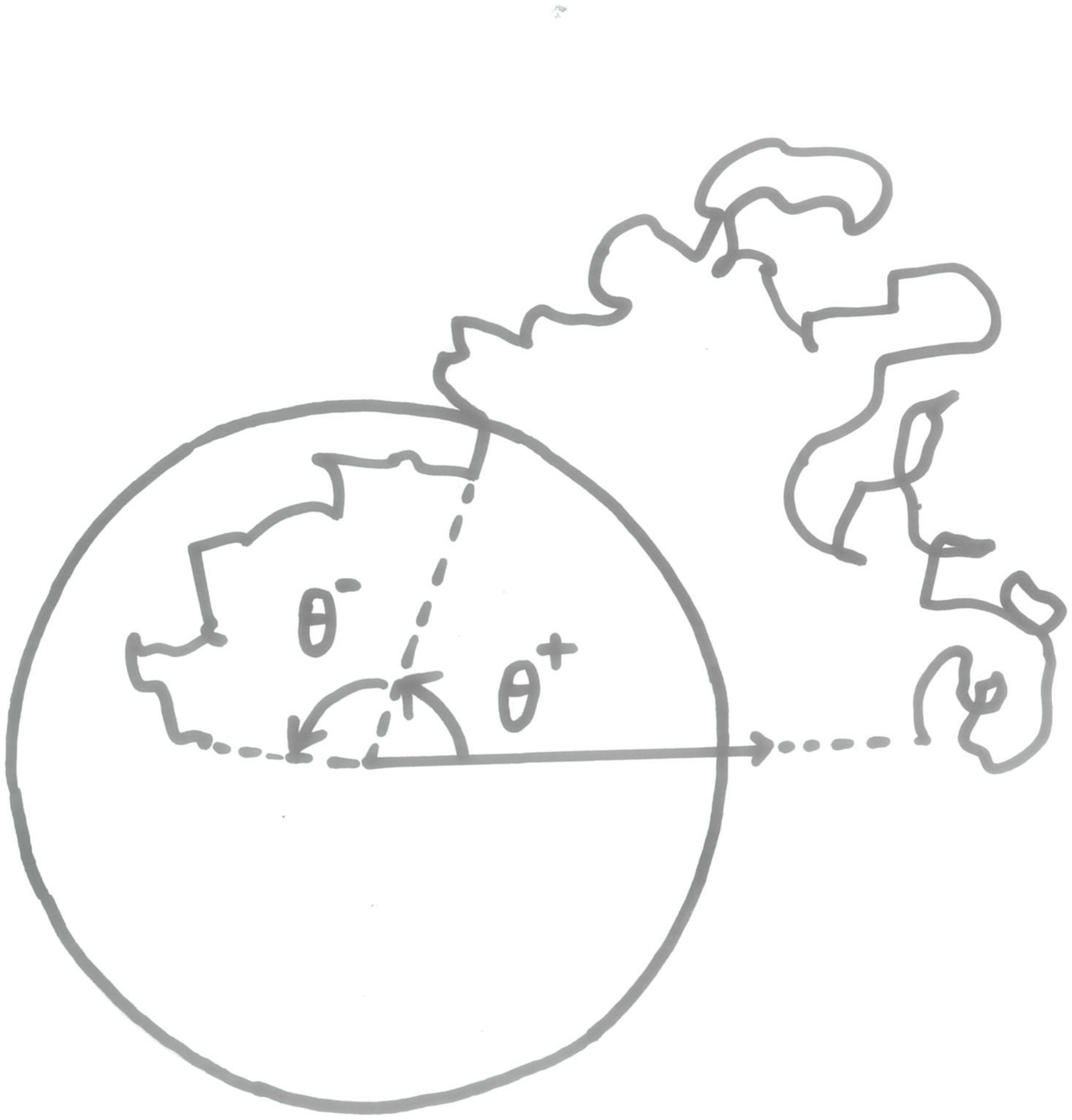}
  \caption{The angles $\theta_+$ et $\theta_-$ for a typical realisation of planar Brownian motion.}
   \label{fig:theta}
\end{figure}

This problem turns out to be closely related to certain questions that arise in polymer physics. Polymers are extended objects which may assume different conformations that affect the medium. In particular, topologically non-trivial configurations in which the polymer has knots or wounds around an obstacle will affect the elastic properties of the medium.
The standard approach is to model the ideal polymer by a random walk on a lattice which, in the continuum limit, 
converges to a Brownian motion.
For planar ideal polymers wound around a point, the problem then reduces to computing the winding number distribution of a Brownian path with fixed ends; Prager \& Frisch \cite{PrFr} and Edwards \cite{Ed}
solved it simultaneously and independently by different methods in 1967.
Edwards' approach, which enables the determination of the elastic properties of ideal polymers in the presence of more general topological constraints, is based on a formal correspondence with the problem of a
quantum mechanical particle in a Bohm--Aharonov magnetic field, and uses path integrals.

A few years later, several authors, Rudnik and Hu \cite{RH} among them, realized that the angular distribution has a non trivial behaviour in the continuum limit: all the moments of $\theta_t$ are divergent.
The origin of this divergence  and, more importantly, the correct way of handling this problem, are in fact discussed in a seminal paper of Messulam and Yor \cite{MY}. In order to probe deeper into the winding process, the winding angle is written as a sum of two terms $\theta_t= \theta^+_t  +\theta^-_t$. Although the terminology had not yet settled at the time, these two terms represent the ``large'' and ``small'' windings respectively. The idea is to describe the winding process as consisting of a series of phases: those in which the particle is far from the origin contribute to  $\theta^+$, whilst those in which the particle makes a lot of turns near the origin bring an important contribution to  $\theta^-$; see Figure 1. 
This decomposition  turned out to be one of the key
concepts enabling the general derivation of the asymptotic laws around several points \cite{PY}; the works that followed, both in the physics \cite{CDM,GF,Sal}  and in the mathematics \cite{Bel} literatures, brought further confirmation of its usefulness. In particular, B\'{e}lisle showed that the asymptotic law for the windings of random walks involves the large-winding component
$$
\frac{2\theta_t^+}{\log t} \,.
$$
This random variable, unlike $2\theta_t^-/{\log t}$,
has moments of all orders.
We emphasise the practical implications of these facts, not only for polymer physics but also in the context  of flux lines in type-two superconductors \cite{DK,Ne}, as well as in the more exotic context of magnetic lines in the solar corona \cite{BR}. 

The study of winding properties in relation to experimental and theoretical work on DNA elasticity has seen a revival in recent years. Although the description of the elastic properties of a single supercoiled DNA molecule requires more complicated models, the short-distance behaviour is similar to the familiar Brownian case \cite{BM}.
Marc Yor, with  D. Holcman and S. Vakeroudis, returned to this theme recently and sought to apply these ideas
to a polymer model inspired by biophysics \cite{VYH}. In that work, the polymer is modelled 
as a collection of $n$ planar rods attached to a point--- each rod making a Brownian angle $\theta_k (t), 1\le k\le n$ with a fixed direction. Although the model does not take into account exclusion constraints, it is sufficiently rich and non trivial to be worth investigating. The object is to compute the mean rotation time ---which is the characteristic time for a polymer to wind around a point:
$$
\tau_n= \inf \{{t>0 :\, \varphi_n(t)=\vert2\pi\vert }\}
$$
where $\varphi_n (t)= \arg Z_n(t)$ is the angle swept by the last rod. 
This reduces to the study of a sum of independent, identically-distributed variables on the complex unit circle with one-dimensional Brownian motions as arguments.  An asymptotic formula for the mean rotation
time $\mathbb{E}(\tau_n)$ in terms of the initial angles of the chain $\theta_k (0)$ is obtained. The physical interest of this work is that it provides an estimate of the mean rotation time  as a function of the number of monomers and  of the diffusion constant.

\section{Brownian windings and the Bohm--Aharonov effect}
\label{bohmSection}
To conclude this discussion we find it particularly instructive to revisit--- albeit briefly--- the link between the windings of Brownian motion and the Bohm-Aharonov effect.

\begin{figure}[!ht]
  \centering
  \includegraphics[scale=0.20]{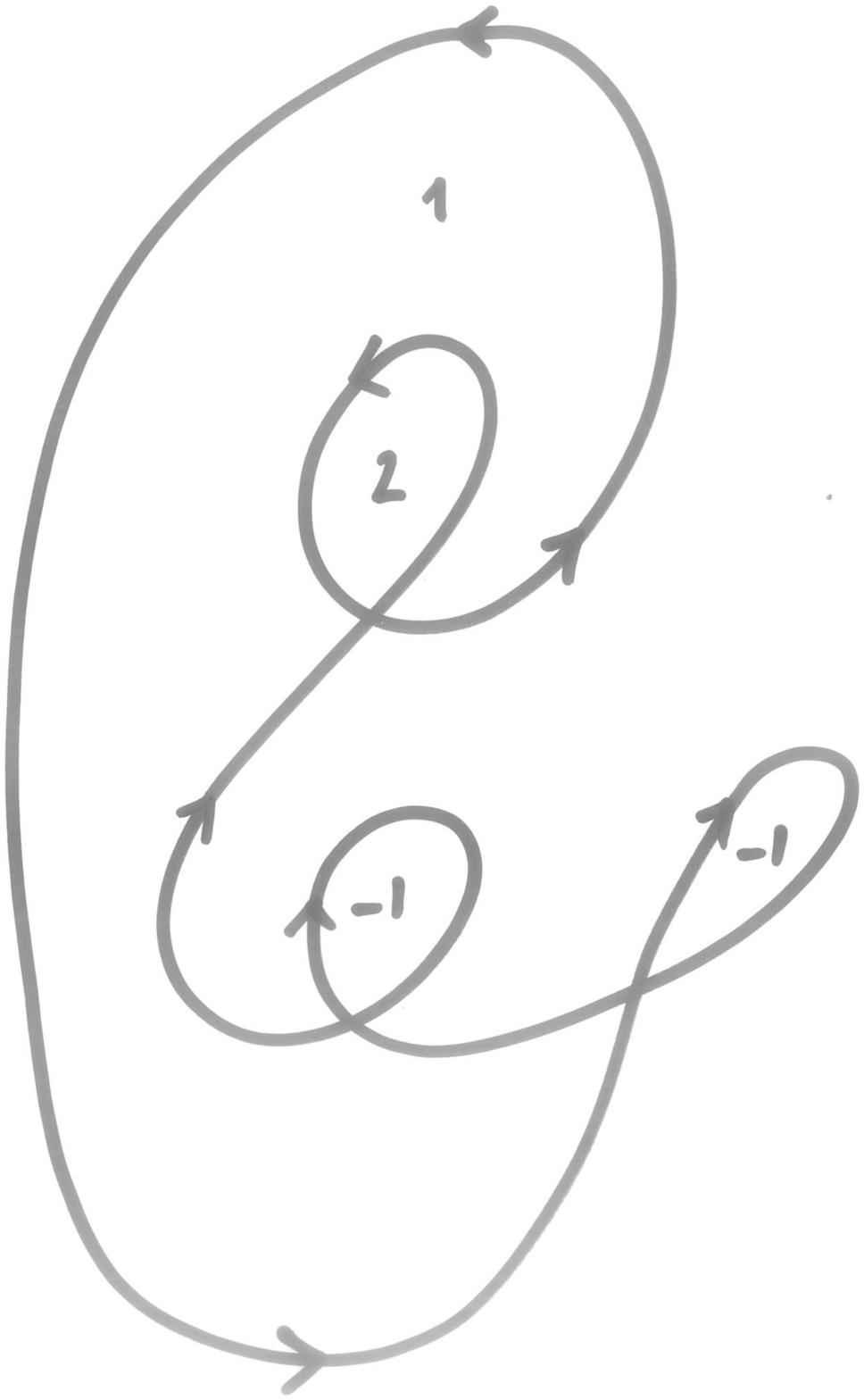}
  \caption{The index $n_z$ of a planar curve with respect to different points  $z$.}
   \label{fig:secteur}
\end{figure}

In this context, the central mathematical object is the index of a curve; its physical counterpart is the notion of magnetic flux. Recall that the index $n_z$ of a 
closed planar curve  $\gamma$ with respect to a point $z$ is an integer giving the number of turns of the curve around that point. Let $A_n$ be the  area of the set of points of a given index $n$. Then the algebraic area of the curve may be decomposed into its winding sectors:
$$
A(\gamma)=\int n_z \, \d z= \sum_{n\in Z} n \,A_n\,.
$$
It should be pointed out  that the  $A_n$  are non trivial random variables which depend on the whole history of the curve. Nevertheless, when the curve is a two-dimensional Brownian bridge,
their expectation may, for $n \ne 0$, be found explicitly \cite{CDO}:
$$
\mathbb{E}( A_n)=\frac{t}{2\pi n^2}\,, \;\; n \ne 0\,.
$$
The idea is to relate these quantities to the quantum partition function  of a charged particle coupled to a Bohm-Aharonov flux  carrying a magnetic flux $\phi=2\pi\alpha $. They are in fact the Fourier coefficients of the difference of two partition functions :
$$
Z_{\alpha}(t)-Z_0(t)= \frac{1}{2\pi t}\sum_{n\in\mathbb{Z}}(\e^{-2i\pi\alpha n}-1)\mathbb{E}(A_n)
$$
where $Z_0$ is the ``free'' partition function, and $Z_\alpha$, $\alpha \ne 0$, is the ``interacting'' one.
It is, however, not so easy to give a precise meaning to these objects. Indeed when $\gamma$ is not a smooth curve,  $n_z$  is not necessarily locally integrable owing to the small turns that the particle makes around the point $z$. W. Werner \cite{We} has shown that there exists a regularized version $n_z^{\epsilon}$ of the index  such that 
$$
A(\gamma)=\lim_{\epsilon\to 0}\int n_z^{\epsilon} \d z
$$
provides a rigorous decomposition of the L\'evy stochastic area into its different components  $n\in\mathbb{Z} \backslash \{0\}$. Note that the stochastic area $A$ does not involve the  $n=0$ sector, in contrast with the arithmetic area ${\mathcal A}$, which includes every sector inside the Brownian bridge:
$$
\mathcal{A}(\gamma)= \sum_{n\in\mathbb Z} A_n\,.
$$

A few years ago, Garban and Trujillo--Ferreras \cite{GT} managed to compute exactly the expected value of the arithmetic area by using LSE techniques :
$$
\mathbb{E}(\mathcal{A})=\frac{\pi t}{5}\,.
$$
Then, by using Yor's result of 1980 \cite{Yo}, they also obtained rigorously the expected values of the area of the $n\neq 0$ sectors,
and thence deduced
$$
\mathbb{E}(A_{0})=\frac{\pi t}{5}-\sum_{n\in\mathbb{Z}} \frac{t}{2\pi n^2}=\frac{\pi t}{30}\,.
$$
The exploitation of these ideas in other physical problems, such as that of determining the density of states of a quantum particle in a random magnetic field, has led  to several more papers
\cite{DFO,GWS}--- thus providing yet another illustration of Wigner's famous observation concerning ``the unreasonable effectiveness of mathematics in the natural
sciences''.

\section{Exponentials of L\'{e}vy processes}
\label{levySection}

We now turn to our second theme; Marc's interest in this topic had its origin in mathematical finance \cite{Yo2}. The following identity, discovered by Dufresne \cite{Du}
in that context, provides a good example of the kind of result which was sure to draw his attention: for $\mu >0$, and $B_t$ a standard Brownian motion,
\begin{equation}
\int_0^\infty  \e^{- \mu t - B_t}\,\d t \overset{\text{law}}{=} \frac{2}{\Gamma_{2 \mu}}
\label{dufresneIdentity}
\end{equation}
where $\Gamma_{\mu}$ is a gamma--distributed random variable with parameter $\mu$. This motivates the general
study of integrals of the form
\begin{equation}
I_t :=  \int_0^t \e^{-W(s)} \,\d s
\label{exponentialAtFixedTime}
\end{equation}
where $W$ is a L\'{e}vy process, i.e. a process started at zero, with right-continuous, left-limited paths, and stationary
increments \cite{Be}. 

We recall here for the reader's convenience
that a L\'{e}vy process $W$ is completely characterised by its
L\'{e}vy exponent $\Lambda$, defined via
$$
{\mathbb E} \left [ \e^{\text{i} \theta W(t)} \right ] = \e^{t \Lambda (\theta)}\,,
$$
and that $\Lambda$ is necessarily of the form
\begin{equation}
\Lambda (\theta) = \text{i} a \theta - \frac{\sigma^2}{2} \theta^2 + \int_{{\mathbb R}_\ast} \left ( \e^{\text{i} \theta y} 
-1 - \frac{\text{i} \theta y}{1+y^2} \right ) \Pi (\d y)
\label{levyKhintchineFormula}
\end{equation}
for some real numbers $a$ and $\sigma$, and some measure $\Pi$ on ${\mathbb R}_\ast := {\mathbb R} \backslash \{0\}$
satisfying
$$
\int_{{\mathbb R}_\ast} \left ( 1 \land y^2 \right ) \Pi (\d y) < \infty\,.
$$
The case $\Pi \equiv 0$ corresponds to Brownian motion with drift, i.e. $W(t) = a t + \sigma B_t$.
The case $\sigma =0$ and $\Pi$ a finite measure yields a compound Poisson process with drift; the
drift coefficient is given by
\begin{equation}
\mu := a - \int_{{\mathbb R}_\ast} \frac{y}{1+y^2} \Pi (\d y)\,,
\label{driftCoefficient}
\end{equation}
the intensity of the process is $\Pi ({\mathbb R}_\ast)$
and the probability distribution of the jumps is
$$
\left ( \Pi ({\mathbb R}_\ast) \right )^{-1} \Pi (\d y)\,.
$$
An important subset of the L\'{e}vy processes for which the theory of the integral $I_t$ takes on a particularly elegant form
consists of the {\em subordinators}. By definition, a subordinator is a non-decreasing L\'{e}vy process; hence $\sigma=0$, the
support of $\Pi$ is contained in $(0,\infty)$, $$
\int_{(0,\infty)} \left ( 1 \land y \right ) \Pi (\d y) < \infty\,,
$$
and $\mu$, as defined in Equation (\ref{driftCoefficient}), is non-negative.
The importance of this class stems from the fact that the inverse local time
of a Feller diffusion is a subordinator.

Yor's work in this area displayed a characteristic fondness for illustrating the theory with concrete
calculations.
Our purpose in what follows is to review some of the techniques and ideas, and to show how they can be adapted to the study of some disordered systems to yield new results in that field.

\section{The disordered system}
\label{disorderedSection}
Mathematically speaking, by ``disordered system'' we simply mean an operator-valued random
variable. The motivations for considering such systems are many, but for physicists the introduction
of randomness is, roughly speaking,  a means of modelling a very complex phenomenon.   
Their study was initiated  by Dyson \cite{Dy}, and gained prominence when P. W. Anderson used a linear difference
equation with random coefficients to explain why the presence of impurities in metals 
has a dramatic effect on their conduction properties \cite{An}. This is one instance of a general phenomenon, now known as {\em Anderson localisation}, whereby linear waves propagating in a medium with randomness tend to have a spatial support
that is localised rather than extended. This phenomenon has since been intensively studied by mathematicians 
and physicists, particularly in the context
of linear second-order differential or difference operators, where the effect of randomness on the spectrum has been the
focus of attention \cite{PaFi}. The main challenges in this respect are to determine conditions under which localisation takes
place and, when it does, to quantify the typical localisation length. The latter problem--- which in many cases
is tantamount to computing the growth rate of a certain subadditive process--- is notoriously difficult \cite{BL, CTT2,Ki}; in what follows, we shall see how Marc's ideas suggested new cases where calculations proved possible.

The connection between
exponential functionals and disordered systems is through their respective applications to diffusion in a random
environment. Indeed, Bouchaud {\em et al.} studied the return probability of a particle diffusing in such an environment
by considering a certain quantum mechanical disordered system  \cite{BCGL}, while Carmona, Petit and Yor 
showed how the moments of $I_t$ yield information on the hitting time of the diffusing particle \cite{CPY}.

Let us describe informally the particular disordered system which will occupy us in the remainder. Consider a diffusion $X(t)$, started from zero, with infinitesimal generator
$$
\frac{1}{2} \e^{W(x)} \frac{\d}{\d x} \left [ \e^{-W(x)} \frac{\d}{\d x} \right ]\,.
$$
The Laplace transform, say $u(x,\lambda)$, of the distribution of the hitting time 
$$
T_x := \inf \left \{ t:\, X(t) = x \right \}
$$
solves the equation
$$
\frac{1}{2} \e^{W(x)} \frac{\d}{\d x} \left [ \e^{-W(x)} \frac{\d u}{\d x} \right ] = \lambda u\,.
$$
If we now set $\psi := \e^{-W/2} u$ and $E=-2 \lambda$, then $\psi$ solves (formally)
the Schr\"{o}dinger equation
\begin{equation}
  - \psi'' + V(x) \,\psi = E \psi
  \label{schroedingerEquation}
\end{equation}
where
\begin{equation}
V(x) := \frac{w^2(x)}{4} - \frac{w'(x)}{2}\,, \;\;\;\; w(x) := W'(x)
\label{supersymmetricPotential}
\end{equation}
and the prime symbol indicates differentiation with respect to $x$.

Of particular interest is the {\em complex Lyapunov exponent} \cite{Lu}
\begin{equation*}
  \Omega (E) := \lim_{x \rightarrow \infty} \frac{\ln \psi (x,E)}{x}
\end{equation*}
where $\psi(\cdot,E)$ is the particular solution of Equation (\ref{schroedingerEquation}) satisfying $\psi(0,E) = 0$ and $\psi'(0,E)=1$.
When $W$ is a L\'evy process, the limit on the right-hand side is a self-averaging (non-random) quantity, i.e. its
value is the same for almost every realisation of $W$. Furthermore, we have
\begin{equation}
\Omega(E) = \gamma(E)  - \i \pi N(E)
\label{complexLyapunovExponent}
\end{equation}
where the real numbers $\gamma(E)$ and $N(E)$ are, respectively, the reciprocal of the localisation length of the
disordered system and the integrated density of states per unit length. 

The recipe for computing the complex Lyapunov exponent is as follows \cite{CTT,FL,Ko}: rewrite the Schr\"{o}dinger
equation with the supersymmetric potential (\ref{supersymmetricPotential})  as the first-order system
\begin{align}
\label{dirac1}
- \psi' - \frac{w}{2} \,\psi &= \sqrt{E} \,\phi \\
\label{dirac2}
  \phi' -  \frac{w}{2} \,\phi &= \sqrt{E} \,\psi 
\end{align}
and introduce the Riccati variable
\begin{equation}
Z := \frac{-1}{\sqrt{E}} \frac{\psi}{\phi}\,.
\label{riccatiVariable}
\end{equation}
Then
\begin{equation}
Z' =  1+ E \,Z^2 - w Z
\label{riccatiEquation}
\end{equation}
and, for $E<0$,
\begin{equation}
\Omega (E) = \frac{c(0)}{2} - E \,{\mathbb E}  \left ( Z_{\infty} \right )\,.
\label{CTTequation}
\end{equation}
In this expression, $Z_{\infty}$ is the unique positive stationary solution of the Riccati equation (\ref{riccatiEquation}),
the coefficient $c(0)$ is the limit as $s \rightarrow 0$ of
\begin{equation}
c(s) := - \frac{\Lambda (\text{i} s)}{s}
\label{levyCoefficient}
\end{equation}
and the expectation is over the realisations of the L\'{e}vy process.

\begin{remark}
For a L\'{e}vy process $W$, the meaning of the stochastic differential equation (\ref{riccatiEquation}) needs to be spelled
out. When $W$ is a Brownian motion, the equation should be understood in the sense of Stratonovich \cite{Ok,RY}.
When $W$ is a compound Poisson process, the equation should be understood as
$$
Z'(x) = 1 + E\,Z^2(x) \quad \text{for $x \ne x_j$}
$$
and 
$$
Z(x_j+) = \exp \left \{ - \left [ W(x_j+)-W(x_j-)\right ] \right \} \,Z(x_j-)
$$
where the $x_j$ are the ``times'' when $W$ jumps; see \cite{CTT,CTT3}.
Hence
the equation makes sense when $W$ is an interlacing process, i.e. the sum of a Brownian motion and of a compound Poisson
process. More general L\'{e}vy processes may be viewed as limits of interlacing processes.
\label{stratonovichRemark}
\end{remark}

Let us now elaborate the relationship between the Riccati variable $Z$ and the exponential $I_t$, defined by Equation (\ref{exponentialAtFixedTime}). For $E=0$, the Riccati equation (\ref{riccatiEquation}) becomes linear and,
by using an integrating factor, 
we obtain
\begin{multline}
Z(x) = Z(0) \,\e^{-W(x)} + \int_0^x \e^{-\left [ W(x)-W(y) \right ]}\,\d y \\
\overset{\text{law}}{=} Z(0) \,\e^{-W(x)} + \int_0^x \e^{-W(x-y)}\,\d y \\
= Z(0) \,\e^{-W(x)} + \int_0^x \e^{-W(s)}\,\d s\,. 
\notag
\end{multline}
In particular, if we suppose that $Z(0)=0$, then we see that the zero-energy Riccati variable has the same law
as $I_x$. So the problem of computing the complex Lyapunov exponent of the disordered system
may be viewed as a generalisation of the problem of computing the first moment of $I_{\infty}$.

\section{Explicit formulae for the distribution}
\label{distributionSection}
One approach which Carmona {\em et al.} used to determine the distribution of $I_{\infty}$ consists 
of expressing it as the stationary distribution of a certain generalised Ornstein--Uhlenbeck process. This process
has an infinitesimal generator, and the stationary density therefore solves a forward Kolmogorov
equation involving the adjoint of this generator \cite{CPY,Pa}.

In the context of our disordered system, the counterpart of the generalised Ornstein--Uhlenbeck process is of course the Riccati process, and the forward
Kolmogorov equation for the probability density $f(x,z)$ of the random variable $Z(x)$ is
\begin{multline}
\frac{\partial f}{\partial x} (x,z) = \frac{\partial}{\partial z} \left \{ \left ( az-1-E z^2  \right ) f(x,z) + \frac{\sigma^2}{2} 
z \frac{\partial}{\partial z}  \left [ z f(x,z) \right ] \right . \\
\left . + \int_{{\mathbb R}_\ast} \left [ \int_z^{z \e^{y}} f(x,t)\,\d t - \frac{yz}{1+y^2} f(x,z) \right ] \,\Pi(\d y) \right \}\,.
\label{fokkerPlanckEquation}
\end{multline}
In particular, the stationary density $f_\infty(z)$ of $Z_{\infty}$ solves an equation that generalises 
to the case $E \ne 0$ Equation (2.2) of \cite{CPY}:
 \begin{multline}
\left ( az-1-E z^2  \right ) f_\infty (z) + \frac{\sigma^2}{2} 
z \frac{\partial}{\partial z}  \left [ z f_\infty(z) \right ] \\ + \int_{{\mathbb R}_\ast} \left [ \int_z^{z \e^{y}} f_\infty(t)\,\d t - \frac{yz}{1+y^2} f_\infty(z) \right ] \,\Pi(\d y) = \text{const}\,.
\label{stationaryEquation}
\end{multline}
It may be shown that, for $E \le 0$, the density is supported on $(0,\infty)$ and that
the constant on the right-hand side of this last equation is in fact zero. 
The calculation therefore simplifies if we set, in the first instance, $E = -k^2$, $k$ real. 
The complex Lyapunov exponent $\Omega(E)$ is 
analytic in $E$ except for a branch cut along the positive real axis. Hence its value elsewhere 
in the complex plane may be obtained by analytic continuation.

\begin{example}
The simplest case arises when 
$$
\Pi \equiv 0\,.
$$ 
Then
$$
f_\infty(z) = C(a,\sigma^2,k)\,z^{-2a/\sigma^2 -1}\,\exp \left [ -\frac{2}{\sigma^2}
(k^2 z +1/z) \right ]\,.
$$
For $k=0$, this is Dufresne's result. For $k > 0$,
one deduces from Formula (\ref{CTTequation}) the result of Bouchaud {\em et al.} 
which expresses $\Omega(-k^2)$ in terms
of the MacDonald functions \cite{BCGL}.
\label{bouchaudExample}
\end{example}

When the L\'{e}vy measure $\Pi$ is non-trivial, there is no systematic method for solving the integro-differential
equation (\ref{stationaryEquation}). Nevertheless, in the particular case where the density of $\Pi$ satisfies 
a differential equation with constant coefficients, it is possible to eliminate the integral term
in (\ref{stationaryEquation}) and so reduce it to a purely differential form \cite{GP}.
\begin{example}
Let  $\sigma =0$ and
$$
\Pi (\d y) = p \,q \,\e^{-q y} \,{\mathbf 1}_{(0,\infty)}(y) \,\d y\,.
$$ 
Then one may deduce from Equation (\ref{stationaryEquation}) that
$$
\frac{\d}{\d z} \left [ (\mu z -1 + k^2 z^2) f_\infty (z) \right ] - p f_\infty (z)
- \frac{q}{z}  \left [ (\mu z -1 + k^2 z^2) f_\infty (z) \right ] = 0
$$
where $\mu$ is defined in Equation (\ref{driftCoefficient}). The solution is given by
\begin{equation}
C (\mu,p,q,k)\, z^q \,( z - z_-)^{-\nu-1} (z_+-z)^{\nu-1} \;\;\text{for}\; 0 < z < z_+\,, 
\label{paulsenDistribution}
\end{equation}
where
$$
\nu := \frac{p}{\sqrt{4 k^2+\mu^2}} \;\;\text{and}\;\; z_{\pm} :=  \frac{1}{2 k^2} \left [ -\mu \pm \sqrt{4 k^2 + \mu^2} \right ]\,.
$$
The normalisation constant is
$$
1/C (\mu,p,q,k) = z_+^{q+\nu} |z_-|^{-\nu-1} \text{B} (\nu,q+1) \,{_2}F_1 \left ( \nu+1,q+1; q+\nu+1; z_+/z_-\right )
$$
and it follows easily that the expectation in Formula (\ref{CTTequation}) for the complex Lyapunov exponent 
is a ratio of hypergeometric functions.
This generalises to an arbitrary negative energy Example B of Carmona, Petit \& Yor \cite{CPY},
and to an arbitrary drift the example first discussed in \cite{CTT3}. Yet more examples may be found 
in \cite{CTT2}.
\label{paulsenExample}
\end{example}

\section{Explicit calculation of the moments}
\label{momentSection}

Another approach to finding the distribution of the exponential functional uses the fact that, at least for subordinators,
the positive moments can be computed exactly; see \cite{BY}, Theorem 2. In order to generalise
this result to our
disordered system, we work with  the Mellin transform of the density $f(x,z)$ of the Riccati variable:
\begin{equation}
\widehat{f}(x,s) := {\mathbb E} \left [ Z^s(x) \right ] = \int_0^\infty z^s \,f(x,z)\,\d z\,.
\label{momentDefinition}
\end{equation}
To find an equation for these moments, we multiply the forward Kolmogorov equation 
(\ref{fokkerPlanckEquation}) by $z^s$ and integrate
over $z$. For $E \le 0$ and $s \ge 0$, the result is
\begin{equation}
\frac{\partial \widehat{f}(x,s)}{\partial x} = s\,\left \{ E \widehat{f}(x,s+1) - c(s) \,\widehat{f}(x,s) + \widehat{f} (x,s-1) \right \} 
\label{momentEquation}
\end{equation}
where $c(s)$ was defined in Equation (\ref{levyCoefficient}), and it is assumed that $f(x,z)$ decays sufficiently quickly at infinity; see \cite{CTT}. To proceed, we introduce the Laplace transform
\begin{equation}
\widehat{F}(\lambda,s) := \int_0^\infty \e^{-\lambda x} \widehat{f}(x,s)\,\d x\,.
\label{laplaceTransform}
\end{equation}
Then
$$
\lambda \widehat{F}(\lambda,s) - \widehat{f}(0,s) = s\,\left \{ E \widehat{F}(\lambda,s+1) - c(s) \,\widehat{F}(\lambda,s) + \widehat{F} (\lambda,s-1) \right \} \,.
$$
In particular, if $E=0$ and the Riccati variable starts at zero, this reduces to the first-order recurrence relation
$$
\widehat{F}(\lambda,s) = \frac{s}{\lambda + s \,c(s)} \widehat{F}(\lambda,s-1)\,,
$$
which is precisely the conclusion of the aforementioned theorem. 

The stationary version of Equation (\ref{momentEquation}) is of course simpler:
\begin{equation}
E \widehat{f}_\infty(s+1) - c(s) \,\widehat{f}_\infty(s) + \widehat{f}_\infty (s-1) = 0\,.
\label{stationaryMomentEquation}
\end{equation}
It has particularly interesting consequences in the subordinator case, where it may be shown that the complex 
Lyapunov exponent has the continued fraction expansion
$$
\Omega (E) = \frac{c(0)}{2} + \cfrac{-E}{c(1) + \cfrac{-E}{c(2)+\cdots}}\,.
$$
Comtet {\em et al.} \cite{CTT} used this in order to determine the low-energy behaviour of the density of states.

Let us end by making it clear that Marc's work on exponential functionals was of course much more than
a collection of striking formulae. With his prodigious knowledge of probability,  he was able to use
exponential functionals as a tool to re-derive or extend results 
obtained by other methods in other parts of the theory of stochastic processes. 
In the particular case where the L\'{e}vy process is a Brownian motion
with drift, his work with H. Matsumoto led among other things to extensions of Bougerol's identity \cite{Bou} and of Pitman's $2M-X$
theorem \cite{BO,BDY,MY1,MY2,Oc}. For more general L\'{e}vy processes, a dominant theme was the
correspondence between  exponential functionals and the semi-stable processes introduced
by J. Lamperti \cite{La}.  There, the main applications are, on the one hand, to the description of the entrance law of the  
semi-stable process and, on the other hand, to the factorisations of the exponential variable \cite{BY,HY}.

\begin{acknowledgement}
It is a pleasure to thank our colleague Christophe Texier for commenting on the manuscript.
\end{acknowledgement}

%
%
%
\biblstarthook{}

\end{document}